# Direction-dependent photo-voltage detection in multifunctional ZnO micro rod/PBTTT-C14 polymer sensor due to gold nanoparticles

Rehan Ahmed and Pramod Kumar*

Department of Physics, Indian Institute of Technology Bombay, Mumbai 400076, India

*Author to whom any correspondence should be addressed

*Email Address: pramod_k@iitb.ac.in

**Abstract**

A sensor that can detect the direction of the incoming light plays a crucial role in further enhancing the versatility of the multifunction sensors for future applications where the sensor can read multiple pieces of information, similar to the biological senses, like skin. A hybrid sensor based on an n-type ZnO micro-rod with p-type optically active organic polymer (PBTTT-C14) is developed for low-cost, large-area piezoelectric and optical sensing applications for future artificial electronic skin. The multi-functionality of the device is achieved due to the heterostructure configuration of vertically aligned piezoelectric ZnO micro rod arrays and PBTTT-C14 polymer between two gold electrodes. The deposition of the top gold electrode also led to the formation of two regions where it forms a continuous film and isolated gold particles (Au NPs). The isolated NPs when activated, has shown surface plasmon resonance (SPR) and Förster resonance energy transfer (FRET) which generate a potential opposite to the normal working of the device, depending on the number of excited Au NPs by the incident light. The polarity flipping/opposite potential development can be attributed to the rise in electron density near top Au contact due to the SPR and FRET mechanism of isolated Au NPs over the PBTTT-C14 which depends on the illumination direction. As a result, direction-dependent photo voltage polarity flipping was realized in the device. The device has produced piezoelectric and direction-dependent photovoltage flipping responses, leading the way for a multifunction sensor that can detect the direction of incident light and touch.

# 1. Introduction

In recent years, advancements in the field of machine learning and robotics have required the development of sensors and devices that can mimic biological senses like touch, smell, vision, hearing, and brain function [1], [2], [3], [4]. The biological sense, like skin, can not only detect touch but also temperature changes; it has multiple other functions like regulating the temperature by sweating, acting as a barrier against harmful substances and pathogens, production of vitamin D, and storage of fat and other nutrients [5], [6]. Taking a cue from the biological senses, it is also logical to fabricate sensors and devices which can perform multiple functions [7], [8]. Inorganic and organic semiconductors have shown photo-detection capability, which can be enhanced by using heterojunctions, including inorganic/organic heterojunctions [9]. Organic semiconductors are carbon-based molecules or polymers and have received a great deal of interest due to their advantages, including solution processing, mechanical flexibility, lightweight nature, low-cost production, and tunable bandgaps [9]. The photo sensors based on the organic semiconductors that harvest the light signals into electrical signals have a wide range of applications in optical communication, health monitoring, and image sensors [10], [11], [12]. The organic materials can be easily processed into the device structure by spin coating and inkjet printing techniques [13]. They are capable of detecting the ultraviolet (UV), visible, and near-infrared (NIR) wavelengths, depending on the bandgap of the material. Inorganic semiconductor-based devices/photodetectors are generally made from materials like silicon (Si) and gallium arsenide (GaAs) which have been extensively used [14], [15]. However, the inorganic thin film devices are relatively difficult and costly to fabricate on flexible substrates and therefore difficult to use in flexible electronic applications, as they have high brittleness and complicated manufacturing process. The majority of inorganic photoactive materials exhibit a fixed light absorption with narrow spectral sensitivity, resulting in the detection of only specific wavelengths within a narrow band [16]. The traditional inorganic materials-based devices due to the above reasons are unsuitable for low-cost flexible photodetectors. Hence, new materials and device heterostructure are needed for enhancement of the device sensitivity and capability. Therefore, the combination of hybrid organic-inorganic heterostructure (O-I HJ) may be considered as excellent device structure for making it highly efficient and more capable. The O-I HJ provides the benefits of both organic and inorganic materials and gives a broader platform for applications in multifunction devices. Several p-type organic semiconductors and n-type inorganic semiconductors from group III-V or II-VI offer an excellent platform for creating p-n junctions [12]. There are several useful inorganic semiconductors such as ZnO and GaN. Interfaces involving inorganic semiconductors with π-conjugated organic semiconductors (polymers) can be utilized for hybrid junctions [17], [18], [19]. The ZnO has advantages over other inorganic semiconductors because of its non-toxic nature and biocompatibility, and it can also be easily synthesized with a low-cost hydrothermal method at low temperatures [20], [21], [22]. Due to its distinct physical properties, including mechanical flexibility and high sensitivity, the ZnO based hybrid organic-inorganic interfaces have attracted significant attention [19]. Various photoactive organic materials such as PBTTT-C14 Poly[2,2'-(1,4-phenylene)-6,6'-bis(4-phenylquinoline)], P3HT (poly(3-hexylthiophene)), PTCDI (Perylenetetracarboxylic diimide), and PEDOT:PSS (Poly (3,4-ethylenedioxythiophene) polystyrene sulfonate) are commonly integrated with the inorganic materials in hybrid photodetector fabrications [23], [24], [25], [26]. Among these materials, PBTTT-C14 is well known for its high charge carrier mobility and stability [27]. PBTTT-C14 exhibits a strong absorption in the UV and visible range, and this absorption profile makes it a suitable and promising candidate for light-harvesting applications in organic photovoltaics [25]. PBTTT-C14 is frequently used as an electron donor (p-type semiconducting material) in organic-inorganic photovoltaic applications. These properties of PBTTT-C14 make it suitable for applications in organic electronics [28].

In this research work, we have demonstrated an integration of hydrothermally grown piezoelectric/semiconductor ZnO micro rods with organic semiconductor PBTTT-C14 polymer. The hybrid ZnO/PBTTT-C14 heterostructure provides both piezoelectric and optical stimulus response, and the directionality of the optical signal is detected with a third component, which is the Au NPs. The electrical signals collected from the multifunction Au NPs/PBTTT-C14/ZnO based hybrid device can have potential applications in piezo-optoelectronics and artificial skin sensors for robotics. The multifunctionality is achieved due to the piezoelectric, charge extraction, and optical transmission nature of ZnO micro rod arrays when it is introduced with an optically active polymer (PBTTT-C14). This heterostructure of ZnO micro rods and PBTTT-C14 is realized for optical and piezoelectric responses without any external voltage, which means that it is a self-powered multifunctional sensor. The ZnO micro rod generates the piezoelectric response and helps in the hybrid charge transfer exciton formation and charge extraction at OI-HJ. The heterostructure of different materials has previously been used for the polarity flipping of the photovoltage for detecting the wavelength of incident light. M. Patel et.al. has shown the polarity flipping of the photovoltage using the p-SnS/p-Si isotype heterojunction for wavelength-selective photodetectors, that exhibit selective photodetection and can detect the UV, visible, and near-infrared (NIR) lights by the polarity of the generated voltage [29]. Similar results corresponding to the particular wavelength-selective detectors with different material heterojunctions are also been realized [30], [31], [32]. Therefore, understanding the charge extraction and energy transfer mechanism in Au NPs/PBTTT-C14/ZnO heterojunction is essential for constructing and describing the device's working mechanism. Surface Plasmon Resonance (SPR) and Fluorescence Resonance Energy Transfer (FRET) have been widely used to study the energy and electron transfer between the ZnO, organic materials, and Au NPs [33], [34], [35], [36]. SPR is an optical phenomenon arising from the collective oscillation of the free conduction electrons on the noble metal nanoparticle's surface when it is excited by photons. Au NPs decorated ZnO surface impacts the system in inducing the accumulation of free electrons and facilitates their transfer across the interface. Recent studies in photoactive organic materials have demonstrated that the FRET mechanism is an effective approach for improving long distance (~1-10 nm) exciton migration [33]. FRET is a non-radiative energy transfer mechanism in which the excitation photon is first absorbed by the donor molecules, and then the energy of the emission of donors is transferred to the acceptors by the interaction of the electric dipoles [37]. In the present work, the competition between the normal operation, which is without much influence of the Au NPs, and with a large influence of Au NPs on the device has been identified as a possible reason for the voltage shift due to the direction of incident light. Further, the study also explores the fluorescence quenching, accumulation of free electrons, and its effect on the direction of electron transport across the interface of Au NPs/PBTTT-C14/ZnO heterojunction. We have shown that the Au NPs/PBTTT-C14/ZnO heterojunction device can detect the directional dependent photovoltage polarity flipping, unlike the previous research where the device is based on wavelength selectivity. This multifunction sensor can be useful for artificial skin in robotics and also for orientation calibration of solar cells to achieve the highest efficiency.

## 2. Experimental details

### 2.1. Material and synthesis process

$SiO_2$/Si was used as the base substrate for the growth of the ZnO micro rod and device fabrication. First, the silicon (Si) wafer was cleaned by the standard RCA cleaning method, and then a 200 nm $SiO_2$ layer was grown on top of the Si by dry thermal oxidation. A bottom electrode was patterned through the photolithography process and Au/Ti (90/10 nm) was deposited on the $SiO_2$/Si substrate by the sputtering method. Hexamethylenetetramine (HMT) ($C_6H_{12}N_4$) and zinc nitrate hexahydrate ($Zn(NO_3)_2 \cdot 6H_2O$) (160 mM solution in 1:1 ratio) were used as the reactant precursor for ZnO micro rods growth via seedless hydrothermal method at 90 °C temperature. HPLC grade DI water (18 MΩ resistivity) was used as a solvent and employed for growth and treatment processes. The reactant precursors were dissolved separately in DI water using a magnetic stirrer, and finally, both were mixed in a closed-lid glass beaker. The Au electrode patterned on the $SiO_2$ substrate ($Au/SiO_2/Si$) was kept face down floating on the solution in a sealed beaker and placed in a preheated oven at 90 °C temperature for 16 hrs, as shown in Figure 1. After completion of growth, the substrate was rinsed thoroughly in warm (50-60 °C) DI water and dried with a nitrogen flow, and then baked on a hot plate for 3-4 min at 100 °C temperature. The detailed process of seedless hydrothermal growth optimization for ZnO growth and its geometry effects is given elsewhere [22], [38]. The PBTTT-C14 solution was prepared with an 8 mg/ml concentration in chlorobenzene and stirred continuously for 50 min at 100 °C temperature in the dark. It was then spin-coated on the ZnO micro rod matrix. All the chemicals were purchased from Sigma Aldrich and used without any further purification. The complete process of the ZnO growth with the hydrothermal method and PBTTT-C14 spin coating is illustrated in Figure 1.

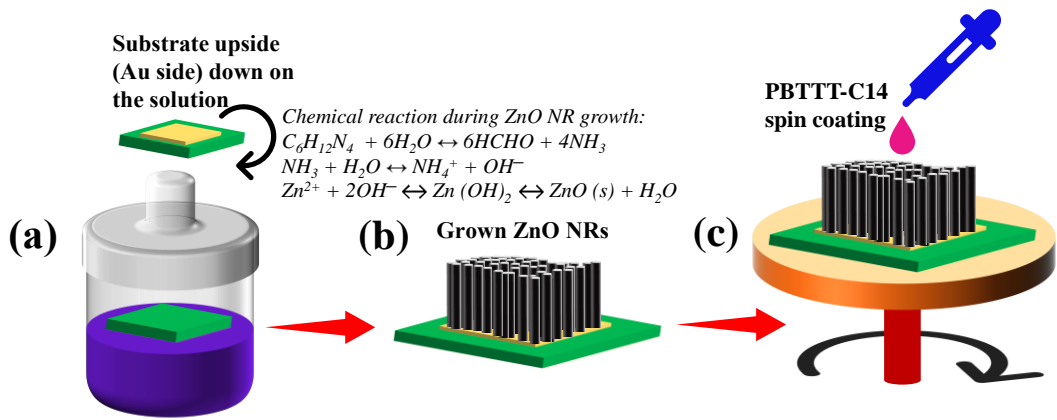

**Figure 1.** Preparation of two precursor concentrations for growth of seedless hydrothermal ZnO micro rods: (a) floating substrate loading onto the growth solution in an enclosed beaker, (b) growth of ZnO micro rods on the substrate, and finally (c) spin coating of PBTTT-C14 on the ZnO micro rod matrix.

## 2.2. Device structure and fabrication

A $SiO_2/Si$ substrate of dimension 7×5 $mm^2$ was employed as the substrate for device fabrication. A thin film of Au (90 nm) as a top electrode was deposited on it the PBTTT-C14 coated ZnO matrix using a metal mask by thermal evaporator which resulted in the formation of Au NPs on some areas of the PBTTT-C14 coated ZnO matrix, especially in the valley regions due to lower Au atom flux. The Au deposited on PBTTT-C14 coated ZnO micro rod shows the formation of isolated Au nanoparticles in the valley regions of the ZnO micro rods.

Device without PBTTT-C14 coating is also prepared with the same fabrication procedure to compare the effect of organic semiconductors. Finally, the whole device structure is packaged with PDMS (polydimethylsiloxane) to provide mechanical support and protect it from moisture and dust after making contacts on the top and bottom electrodes. The schematic architecture of the device is presented in Figure 2, in which the large view of the single ZnO micro rod shows the light penetration into the PBTTT-C14 layer through the ZnO micro rod, where the gold film is non-continuous. The uncoated surface of the micro rod allows the incident photon to enter into the PBTTT-C14 layer, and the process of photon absorption and charge extraction

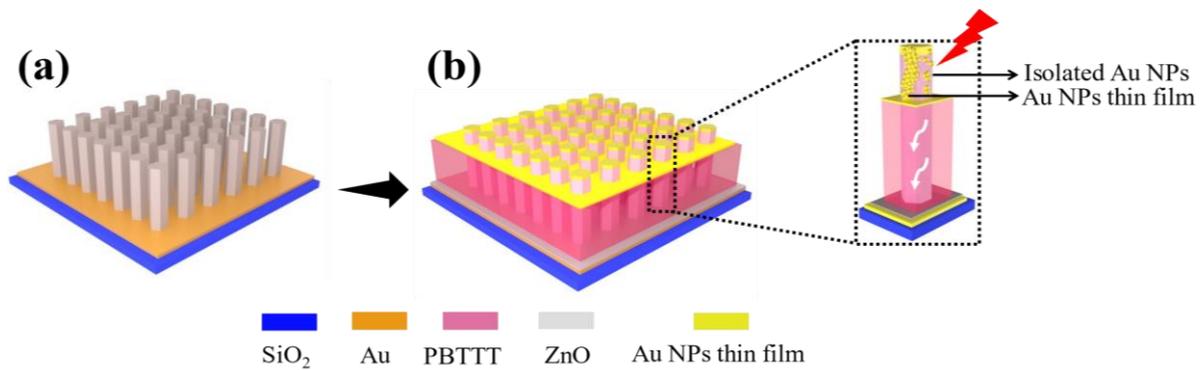

in an Au NPs-O-I heterojunction takes place.

**Figure 2.** Schematic design of the hybrid device structure, (a) represents the growth of ZnO NR arrays and (b) spin coating of PBTTT-C14 and then Au thin film deposition which resulted in the formation of Au NPs. The enlarged figure shows the light penetration through the single ZnO micro rod into the PBTTT-C14 layer.

**2.3. Characterization techniques**

Morphological analysis of the grown ZnO micro rod was conducted with a Field Emission Scanning Electron Microscope (FE-SEM, JSM-7600F). The photoluminescence (PL) measurement was done through the designed setup which consists of the He-Cd laser as the excitation source of wavelength (λ) 325 nm. The source had an energy of 3.81 eV, which is sufficient to excite the electron in the ZnO semiconductor. PerkinElmer Lambda 950 UV-vis spectrophotometer was used for the absorption measurement. The optical measurement of the sensor was characterized by a solar simulator (ORIEL, LSH-7320 ABA LED Solar Simulator) and a semiconductor characterization system (Keithley 4200-SCS). The intensity of the solar simulator was 100 mW/cm$^2$ and is equal to one sun. A weight-adjustable tapping mechanism setup is used to apply force on the device, and the generated piezoelectric voltage is measured with the help of a digital oscilloscope (Tektronix MDO4054-3).

## 3. Results and discussion

SEM images of grown ZnO micro rods with PBTTT-C14 coating and top Au deposition are shown in Figure 3. The top view image in Figure 3 (a) shows ZnO micro rods with spacing and the inset image shows the Au NPs film with size approximately 20-25 nm. The cross-section image in Figure 3 (b) shows that grown ZnO micro rod arrays are vertically aligned along the *c*-axis, dense, and with average diameter and length ranging from 500-700 nm and 5-6 µm, respectively. The image also shows the formation of Au NPs thin film and isolated Au NPs at some portions of the PBTTT-C14 coated ZnO micro rods.

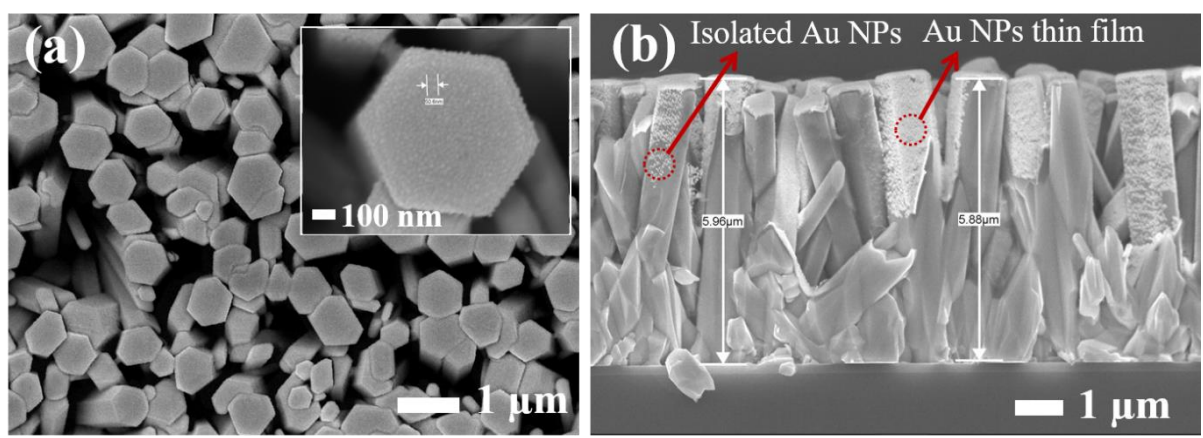

**Figure 3.** SEM images of Au deposited PBTTT-C14-coated ZnO micro rod, (a) top view and the inset shows the high magnification top view image (b) cross-sectional view.

PL and absorption measurements were carried out to investigate the changes in optical properties of PBTTT-C14 thin films and when it was coated on ZnO micro rods based hybrid devices. The PBTTT-C14 was spin coated on a transparent glass slide and an optical scan from 400 to 800 nm was carried out to determine its absorption spectrum. PBTTT-C14 polymer film exhibits strong absorption in the visible range, with absorption maxima at ~ 546 nm, as shown in Figure 4 (a). The observed spectrum shows a broad absorption band between the range 480─650 nm, and the absorption peak is equivalent to 2.27 eV, which is in good agreement with the previously reported result [25]. Figure 4 (b) shows the PL spectrum of the same PBTTT-C14 polymer film when excited with an excitation wavelength of 325 nm. It shows the wavelength range of emission spectra of the PBTTT-C14 thin film. It can be seen from Figure 4 (b) that the PBTTT-C14 thin film shows PL between 600─800 nm, and a maximum emission at 660 nm hence making it suitable for visible photo sensing applications.

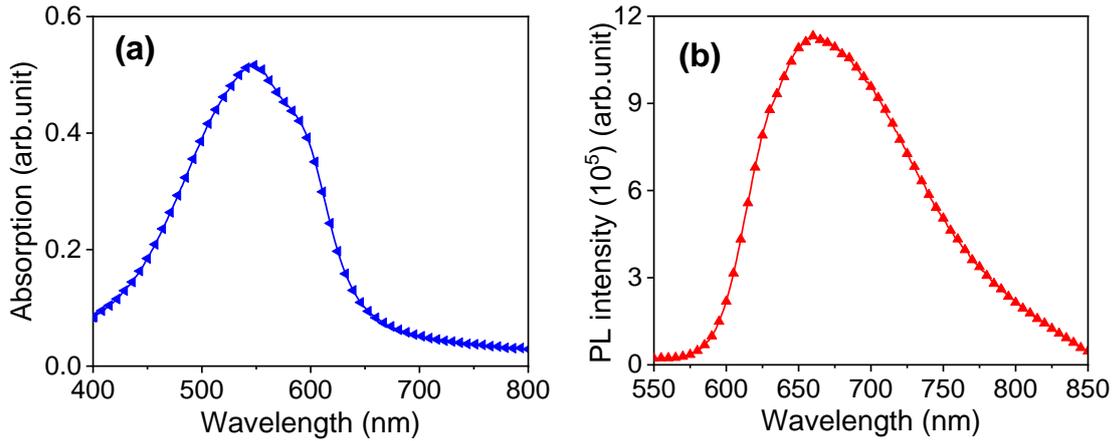

**Figure 4.** The absorption (a) and PL spectra (b) of PBTTT-C14 thin film.

The PL spectra of PBTTT-C14 coated ZnO micro rods and ZnO micro rods were also measured to see the effect of PBTTT-C14 and ZnO interface on the PL spectra. First, the PL spectrum of the ZnO micro rod array was taken, and later the solution of PBTTT-C14 was spin coated on the ZnO micro rod matrix, and PL measurement was again conducted. The PL spectra were compared for both bare ZnO and PBTTT-C14/ZnO to understand the charge carrier transfer, extraction, and recombination of the photo generated electron and holes pairs in hybrid heterojunction. Figure 5 (a) shows the comparison of PL spectra of ZnO micro rod and PBTTT-C14 coated ZnO micro rod at room temperature. The PL spectrum of the ZnO micro rods demonstrates three typical emission bands in which the UV emission band at 378 nm and 381 nm are respectively due to the free and bound exciton recombination and the broad green emission band (576 nm) is due to the intrinsic defects such as zinc interstitial zinc (Zn) and oxygen (O) vacancies [39]. The much lower value of the broad green emission near the wavelength range of 500─600 nm shows that the Zn or O defect vacancies have been significantly reduced. These results indicate the relatively pure crystalline *c*-axis growth of ZnO micro rods. After the PBTTT-C14 coating on the ZnO matrix, the PBTTT-C14/ZnO gives similar PL emission bands to those of ZnO, but the band intensity is quenched, especially in the visible emission, as shown in the insets of Figure 5 (a). The PL quenching indicates the transfer of charge carriers at the interface [40]. The PBTTT-C14 coated ZnO micro rods film shows variation in PL intensity pattern in the 500-600 nm range without much change in the UV peaks except for the decrease in the intensity due to the top PBTTT-C14 layer. The variation in the PL spectrum in the visible range may be due to the charge transfer that occurs between ZnO micro rods and PBTTT-C14 polymer, leading to nonradiative recombination, which can be seen in the insets of Figure 5 (a). The deposition of the top Au electrode is done with a 90 nm film by thermally depositing it on the PBTTT-C14 coated ZnO micro rods. This process leads to the formation of Au NPs and continuous films with NPs in some portions of the ZnO micro rods, where there is a relatively lower flux of Au atoms. The size of Au NPs, ranging from 25-30 nm, is observed on the ZnO micro rods from the SEM images.

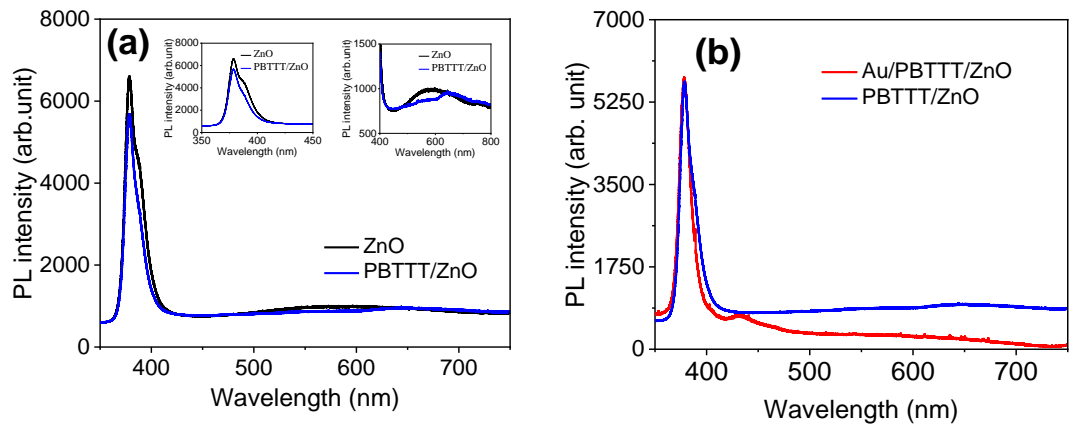

**Figure 5.** PL spectra comparison of PBTTT-C14 polymer spin-coated on ZnO micro rod matrix without (a) and with the deposited Au NPs/film (b). The insets show the intensity quenching of PL spectra.

Figure 5 (b) compares the PL spectrum of the PBTTT-C14/ZnO micro rods heterostructure with and without Au NPs, which shows a quenching in the visible region. A plasmonic peak at 432 nm confirms the size of Au NPs [41]. The quenching shows the effect of Au NPs on the light absorption and charge transfer process due to the surface plasmon resonance energy transfer.

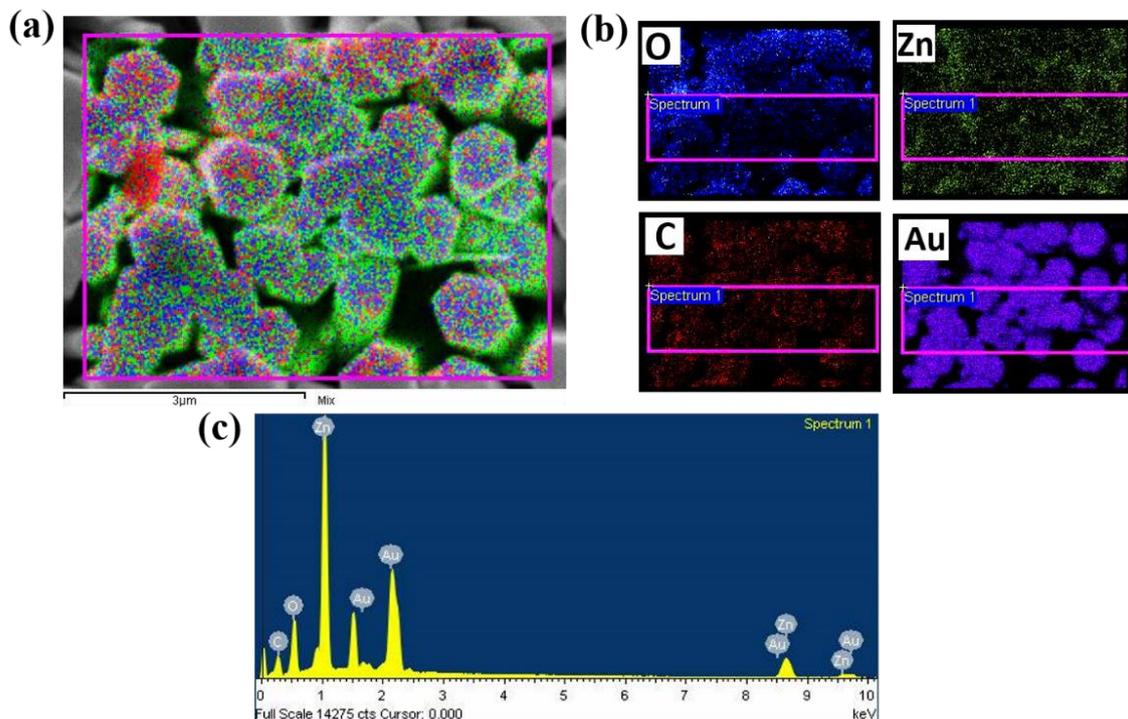

**Figure 6.** Energy dispersive spectroscopy (EDS) (a), elemental mapping analysis (b), and EDS spectrum (c) of Au NPs/PBTTT-C14/ZnO micro rod heterojunction.

Figure 6 shows the energy dispersive spectroscopy (EDS) for quantitative chemical composition analysis of the sample. The elemental mapping of Au NPs/PBTTT-C14/ZnO micro rods heterojunction shows the spatial distribution of elements within the sample. The EDS spectrum shows prominent peaks corresponding to carbon (C), oxygen (O), zinc (Zn), and gold (Au), indicating their presence in significant amounts. Quantitative analysis shows that the sample contains 9 wt% carbon, 14 wt% oxygen, 36 wt% zinc, and 39 wt% gold. The gold levels are much higher as the measurement was performed from the top of the sample. Two devices with and without PBTTT-C14 polymer film were prepared to see the effect of Au NPs on the charge generation and extraction process in Au/PBTTT-C14/ZnO/Au and Au/ZnO/Au devices. The I-V characteristics of the fabricated devices in the dark and at three different incident directions of incoming light are shown in Figure 7. The responses of the devices were measured under the illumination of one sun at a vertical incidence angle (90°) and two tilt modes, approximately at 90±45°. Excitons can be created in both ZnO micro rods and PBTTT-C14 polymer, and both devices show the photovoltaic effect, as the photo-generated electron-hole pairs are separated to generate photovoltage. The discontinuous Au top contact due to the protruding and spaced ZnO micro rods also provides light penetration into the device due to its transparency, which makes it suitable for light penetration into the buried structure, as shown in Figure 2. The fabricated photo sensors are placed directly below the light source at approximately 30 cm, and the measurements are done at different angles/direction (±45°) by tilting the device.

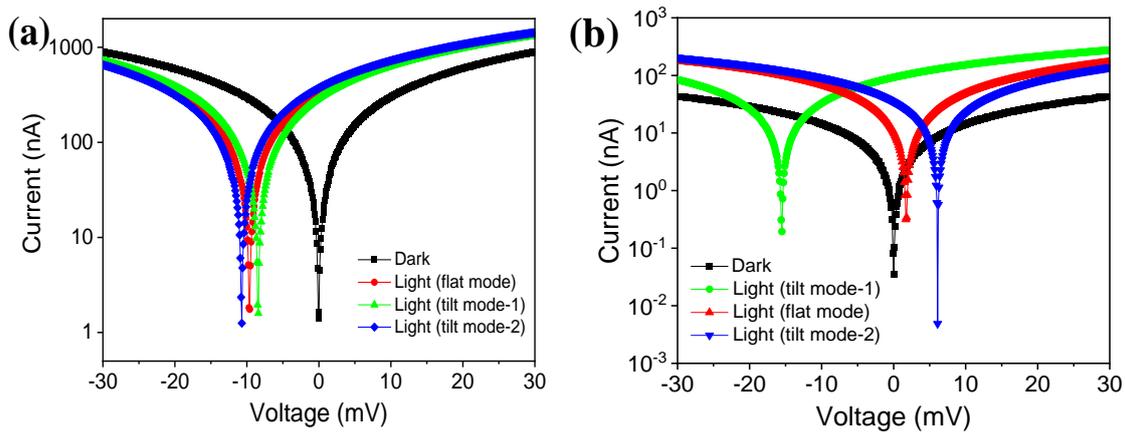

**Figure 7.** I-V characteristics of device performance: (a) Au NPs/ZnO and (b) polarity flipping of photovoltage corresponding to different tilt modes of the device under light illumination in Au NPs/PBTTT-C14/ZnO micro rods. The tilt modes 1 and 2 describe the angle of the incident light illumination with ±45 degrees, and the flat mode is for the normal light incident.

The devices with and without the PBTTT-C14 layer were tested under the same conditions. Figure 7 (a) shows the performance of the device without the PBTTT-C14 layer and the result shows that the generated photovoltage is around -10 mV under normal incidence/flat mode, similar results were also seen in the previous investigations [42], [43]. Changing the angle of incident light in tilt modes 1 and 2 resulted in some minor change in the voltage, which is around ±1 mV. Figure 7(b) shows the results of the device with PBTTT-C14 polymer layer on the ZnO micro rod matrix. The device, under light illumination, when tilted at different directions/angles, shows a large variation in the generated photovoltage. The developed device exhibits a change in photovoltage from positive to negative after crossing the normal incidence of illumination, which can be called the voltage flipping of the device under different illumination directions/angles. Therefore, this novel feature of the device enables it to detect the direction of the incident light. In Figure 7 (b), different tilt modes show different values and signs of the generated voltage. The complete process of photon absorption and its extraction into charges can be understood by following the steps: (i) When photons are absorbed within the organic semiconductor, bound excitons are generated, which then diffuse to the OI-HJ. The diffusion of the generated excitons to the interface produces charge transfer excitons. (ii) At this junction, the electron transfers to the inorganic layer via either a resonant or nonresonant process, depending on the relative energy gaps and offsets between the organic and inorganic materials [44]. The interaction forms a hybrid charge transfer exciton (HCTE), whose stability is mainly influenced by the permittivities of the materials in contact and the effective mass of the electron in the inorganic material. (iii) The HCTE subsequently dissociates, and the resulting electron and hole are collected at the electrodes, returning the organic semiconductor back to its ground state. The bandgap tuning in photo-active semiconductors plays a key role in electron excitation and charge transformation or extraction within or on electrodes. Hence, the efficiency and performance of the photo device can be directly affected by the bandgap tuning [45]. In the inorganic photodetector (without PBTTT-14 in Figure 7 (a)), the photoactive semiconductor ZnO was decorated with Au NPs on its surface, which induced the surface plasmon resonance, enhancing the UV response of ZnO based photodetectors and generating -10 mV photovoltage under light illumination. This effect is due to the defect level emissions in ZnO micro rods, which cause surface plasmon resonance in the Au NPs, enhancing the local electromagnetic field around the Au nanoparticles. Consequently, a significant number of electrons are excited by the Au NPs [42], [34]. When the organic semiconductor (PBTTT-C14) is introduced on the ZnO micro rod arrays, it forms a hybrid O-I heterojunction and the result is remarkably different compared to the normal ZnO device, as seen in Figure 7 (b). The hybrid O-I device exhibits shifting of photovoltage from positive and negative values according to direction of the illumination. The polarity flipping of the photovoltage may be due to the regions in the devices where Au NPs density varies, which play an active role in the charge carrier transport process in PBTTT-C14 and ZnO heterostructure. The shift in the voltage in the Au/ZnO/Au device, which is shown by Figure 7 (a) under the light illumination, can be understood by the energy band diagram of Au NPs/ZnO heterostructure through the SPR, as shown in Figure 8. The Fermi level of ZnO is lower than the Fermi level of Au from the vacuum level, which promotes the flow of electrons from the Au NPs to ZnO micro rods. The direct contact of noble metal and semiconductor forms a new equilibrium of the Fermi level ($E_f$) of the whole system, which is closer to the conduction band of the ZnO, and the Schottky barrier is formed. The PL spectra of ZnO usually consist of a UV emission (near band edge) and possibly one or more visible bands. The UV emission (100—400 nm) is due to the recombination of free excitons while the visible emission (400—800 nm) is mainly responsible for defects, such as oxygen vacancies, zinc vacancies, oxygen interstitials, and zinc interstitials, which are relatively common features for ZnO nanostructures [39]. The surface plasmon in Au NPs can excite the defect level emission energy in the ZnO due to the energy matching between the surface plasmon absorption of the Au NPs and the defect emission in ZnO micro rods [46]. The energy levels of these defects vary from 0.05 eV to 2.8 eV within the bandgap, which gets excited under light illumination [47]. In Figure 5 (b), the Au NPs show the plasmon band at 432 nm (2.87 eV). When the light is illuminated, the photogenerated electrons are excited from the valence band to the conduction band in the ZnO, and at the same time, the different defect levels in the ZnO are also excited. Simultaneously, the SPR oscillations in the Au NPs are excited, these excited resonant electrons are so active that they can elevate electrons from the surface of the Au NPs to the conduction band (C.B.) of the ZnO (route 1 in Figure 8). The resonance energy of the Au NPs (1.77—2.75 eV) is much higher than the Schottky barrier height of 0.6 eV in the Au/ZnO interface, which causes the resonant electron to cross the Schottky barrier and reach the ZnO conduction band. The defect level emissions (deep-level emission) in ZnO are absorbed by the Au NPs (route 2 in Figure 8) through the SPR due to the resonance energy of the Au NPs (450 —700 nm) being close to the emission energy of the ZnO in defect levels (450 —700 nm) [42]. The transferred electrons from the defect states of ZnO to the Au NPs increase the resonant electron density in Au NPs and result in creating energetic electrons in the higher energy state [48]. Thus, the migration of electrons from the surface of the Au NPs has significantly increased the electron density in the conduction band of the ZnO, however, the electron density is slightly decreased on the Au NPs surface [49]. Additionally, another possibility is that the photogenerated holes in the ZnO can be trapped or captured by the Au NPs, effectively suppressing electron-hole recombination in the device. As a result, higher carrier charge density is collected by the electrodes, thereby enhancing the photovoltage under light illumination [43], [50]. Since the deposition of Au NPs on the side wall of ZnO micro rods is not uniform, the SPR resonant electrons may be more active in the higher density of NPs and vice versa, which results the slight shift of the photoresponse in the device at different tilt modes, as shown in Figure 7 (a).

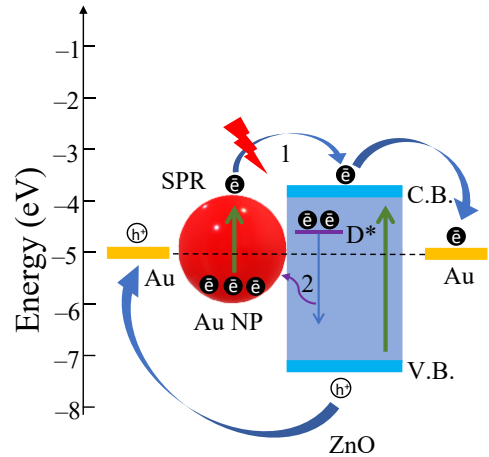

**Figure 8.** Schematic energy band diagram of Au/ZnO/Au device with Au NPs elucidating the process of surface plasmon coupling between Au NPs and ZnO micro rods. Where the C.B., V.B., and D* represent the conduction band, valence band, and defect level in the ZnO, respectively.

The polarity flipping in the Au NPs/PBTTT-C14/ZnO micro rods device, as shown in the result of Figure 7 (b), can be understood by the band level diagram of Figure 9. FRET is a process that can shed light on the charge carrier dynamics over a distance between two different molecules [33], [51], [52]. It is a non-radiative energy transfer mechanism by which the energy (emitted photon) is transferred from one molecule to an excited state to another molecule. In Figure 5 (b), the quenching phenomenon observed in the Au NPs/PBTTT-C14 may result from the interplay between the FRET and SPR effects [49]. The PBTTT-C14 device can work in two modes, one where the density of active Au NPs is negligible which can be called normal device operation and the second where the density of active Au NPs is very large due to the incident light which can be described as active mode. These two competing processes can give rise to a photovoltage polarity flipping phenomenon. The normal mode of device operation is shown in Figure 9 (a) where it is assumed that the density of Au NPs is very less. In this case, PBTTT-C14 absorbs energy and forms excitons which get separated into electron and hole and the energy levels decide the charge carrier flow direction. The electron is collected by the bottom Au electrode and hole by the top Au electrode. The active mode of operation is shown by Figure 9 (b) where the incident light falls on large number of Au NPs. Here SPR and FRET mechanism work together, where Au NPs act as acceptors and the energy from the donor PBTTT-C14 polymer is transferred non-radiatively to the Au NPs. The overall effect of this process results in the reversal of charge carrier flow due to the accumulation of higher electrons close to the top Au electrode, which is shown by Figure 9 (c). Hence, the accumulated electron gets collected at the top Au electrode, and holes move through the ZnO defect level toward the bottom contact, giving rise to photovoltage polarity flipping.

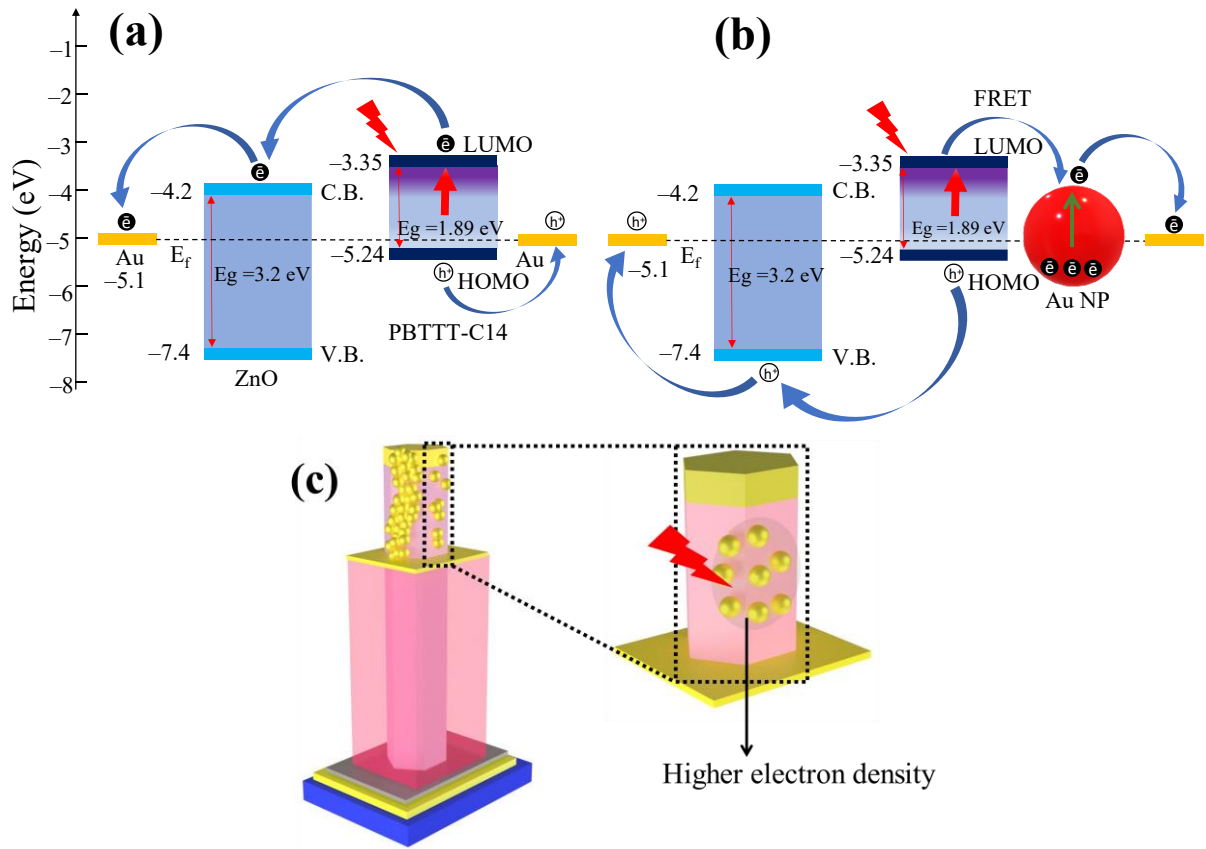

**Figure 9.** Schematic energy band diagram of Au/PBTTT-C14/ZnO/Au device (a) without considering Au NPs (normal mode operation) and (b) with Au NPs (active mode operation), corresponding to different tilt modes of the devices under light illumination, (c) schematics of the process of electron density accumulation near the top Au contact due to the Au NPs.

The piezoelectric response of the device is characterized by a laboratory made weight adjustable tapping setup. A pressure of approximately 35 kPa was applied on the device by using the tapping setup, and the generated output piezoelectric voltage was measured with an oscilloscope. Figure 10 shows the piezoelectric output voltage of around 10 mV while the press and release cycle of the pressure. The positive and negative output voltages are by the press and release cycle of the force. The obtained output voltages while press and release are ~10 mV. Under the applied stress, the compressive deformation in the micro rods is induced symmetrically along the *c*-axis (longitude) of the micro rods. The output piezoelectric voltages while press and release could be attributed to the back-and-forth flow of the electrons in the circuit driven by the piezoelectric potential [53].

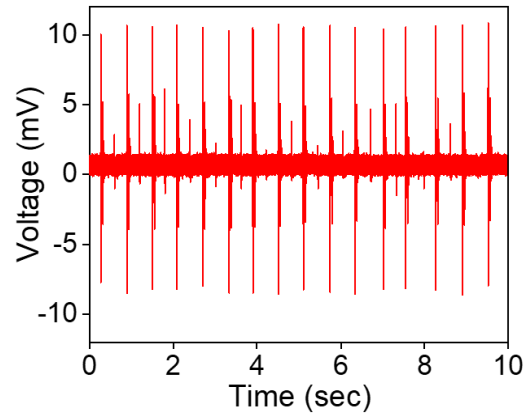

**Figure 10.** The piezoelectric output result of the device while pressing and releasing the pressure.

## 4. Conclusion

Introducing the organic semiconductor (PBTTT-C14) into the ZnO micro rod matrix has shown a photovoltage polarity flipping effect, where the generated photovoltage is shifted from positive to negative values under different illumination directions. The photovoltage polarity flipping can be explained by the number of active Au NPs, which reverses the flow of the charge carriers. When a large number of Au NPs get active due to a certain direction/angle of incident light, both FRET and SPR take place, which results in the accumulation of electrons in the vicinity of the top Au electrode, resulting in photovoltage polarity flipping. The normal mode of device operation takes place at a direction/angle where the Au NPs are relatively less active and electrons are guided by the band levels of PBTTT-C14 and ZnO toward the bottom Au electrode. Hence, the competition between the normal mode working and active mode when SPR and FRET mechanisms of Au NP and PBTTT-C14 are active leads to the development of a light directional detection sensor. The proposed device can also produce approximately ~10 mV piezoelectric signal, making it suitable for multifunctional sensing applications.

## Acknowledgment

The authors would like to thank the Indian Institute of Technology Bombay for the seed grant and the Science and Engineering Research Board (SERB), Government of India for funding. R. Ahmed would like to also thank the Council of Scientific and Industrial Research (CSIR), Government of India for providing the fellowship. The authors would also like to thank Amandeep Kaur and Phalguna Srinivasan for their help in PL and optical measurements.